\documentclass[11pt, times,twocolumn]{article}

\usepackage{graphicx}
\usepackage{subfig}
\usepackage{amssymb}
\usepackage{url}
\usepackage{cite}

\begin{document}
\topmargin 0pt
\oddsidemargin 0mm
\renewcommand{\thefootnote}{\fnsymbol{footnote}}
\begin{titlepage}

\vspace{5mm}

\begin{center}
{\Large \bf {Tailoring the structural and electronic properties of graphene-like ZnS monolayer using biaxial strain} } \\
\vspace{6mm}
{Harihar Behera$^{1, 2, a}$ and Gautam Mukhopadhyay$^{2, b}$} \\
{$^1$ School of Technology, The Glocal University, Delhi-Yamunotri Marg (SH-57),
Mirzapur Pole, Dist.-Saharanpur, U.P. - 247001, India} \\
{$^2$ Department of Physics, Indian Institute of Technology, Powai, Mumbai-400076, India}\\
{$^a$ E-mail: harihar@phy.iitb.ac.in; behera.hh@gmail.com} \\
{$^b$ Corresponding author's E-mail: gmukh@phy.iitb.ac.in; g.mukhopa@gmail.com} \\


\end{center}

\vspace{5mm}
\centerline{\bf {Abstract}}
\vspace{5mm}
 Our First-principles Full-Potential Density Functional Theory (DFT) calculations show that a monolayer of ZnS (ML-ZnS), which is predicted to adopt a graphene-like planar honeycomb structure with a direct band gap, undergoes strain-induced modifications in its structure and band gap when subjected to in-plane homogeneous biaxial strain ($\delta$). ML-ZnS gets buckled for compressive strain greater than 0.92\%; the buckling parameter $\Delta$ (= 0.00 \AA\, for planar ML-ZnS) linearly increases with increasing compressive strain ($\Delta = 0.435$ \AA \,at $\delta = - 5.25$\%). A tensile strain of 2.91\% turns the direct band gap of ML-ZnS into indirect. Within our considered strain values of $|\delta| < 6\%$, the band gap shows linearly decreasing (non-linearly increasing as well as decreasing) variation with tensile (compressive) strain. These predictions may be exploited in future for potential applications in strain sensors and other nano-devices such as the nano-electromechanical systems (NEMS).\\

\end{titlepage}

\section{Introduction}
The study of 2D crystals is an emerging field of research
  inspired by the recent phenomenal growth in the research on graphene
  (a one atom-thick nanocrystal of carbon atoms tightly-bound by the
  sp$^2$-hybridized bonds in a 2D hexagonal lattice) which promises many novel applications \cite{1, 2, 3}. Representative samples of some 2D/quasi-2D nanocrystals that have been synthesized recently include BN \cite{1, 2, 3}, MoS$_2$ \cite{1, 2, 3}, MoSe$_2$ \cite{1, 2, 3}, Bi$_2$Te$_3$ \cite{1, 2, 3}, Si \cite{4}, ZnO \cite {5}. In 2006, using density functional theory (DFT) calculations, Freeman et al. \cite{6} predicted that when the layer number of (0001)-oriented wurtzite (WZ) materials (e.g., AlN, BeO, GaN, SiC, ZnO and ZnS) is small, the WZ structures transform into a new form of stable hexagonal BN-like structure. In 2007, this prediction was experimentally confirmed in respect of ZnO \cite{5} whose stability is attributed to the strong in-plane sp$^2$-hybridized bonds between Zn and O atoms. ZnS and ZnO have similar atomic structures and comparable chemical properties \cite{7, 8}. ZnS is a direct band gap (E$_g = 3.72$ eV in normal cubic zinc blend phase and E$_g = 3.77$ eV in hexagonal wurtzite phase) semiconductor which promises a variety of novel applications as light-emitting diodes (LEDs), electroluminescence, flat panel displays, infrared windows, photonics, lasers, sensors, and bio-devices etc. \cite{7, 8}. Further, due to its larger band gap than that of ZnO (E$_g$= 3.4 eV), ZnS is more suitable for visible-blind ultraviolet (UV)-light sensors/photodetectors than ZnO \cite{5}. ZnS nanostructures in the form of nanoclusters, nanosheets, nanowires, nanotubes and nanobelts are currently being intensely studied both experimentally and theoretically \cite{7, 8}. However, the study of a monolayer graphene (MLG) analogue of ZnS (ML-ZnS) is rare \cite{9}, wherein the authors use the plane wave and projected augmented wave-methods for their calculations and the main result is that a highly reconstructed surface with its energy 0.08 eV/ZnS lower than that of a perfectly planar structure is the most stable structure of ML-ZnS. Here, we report our DFT calculations on the effect of hexagonal symmetry preserving homogeneous biaxial strain on the structural and electronic properties of ML-ZnS. Our study simulates an ideal experimental situation in which ML-ZnS is supported on a flat stretchable substrate. The main advances of our study include
  \begin{enumerate}
\item [i.]{the establishment of the graphene-like planar stable structure of ML-ZnS (see Figure 1) based on all electron full-potential DFT calculations (which is different from the result of \cite{9} based on mechanical annealing followed by microcanonical molecular dynamics with classical interatomic potentials); }


\item[ii.]{the determination of the minimum strain beyond which the flat structure would acquire a buckled structure characterized by a parameter called bucking parameter $\Delta$ (see the caption of Figure 1 for an explanation $\Delta$);}
  \item[iii.]{the prediction of a linear variation of $\Delta$ with the in-plane lattice parameter $a$ (or strain) within certain range of $a$ values;}
  \item[iv.]{the calculation of the 2D bulk modulus of ML-ZnS and its comparison with that of graphene and ML-BN;}
  \item[v.]{the prediction of a direct-to-indirect gap-phase transition at certain tensile strain value within the interval (2.777\%, 2.908\%)};
  \item[vi.]{the prediction of the nature of variation (linear or non-linear) of band-gap with equi-biaxial strain}.
  \end{enumerate}
To our knowledge the above items (i-vi) are new results, which are based on the all electron full-potential DFT calculations with two atoms per unit cell.
\begin{figure}[h]
 \centering
 \includegraphics[scale=0.9]{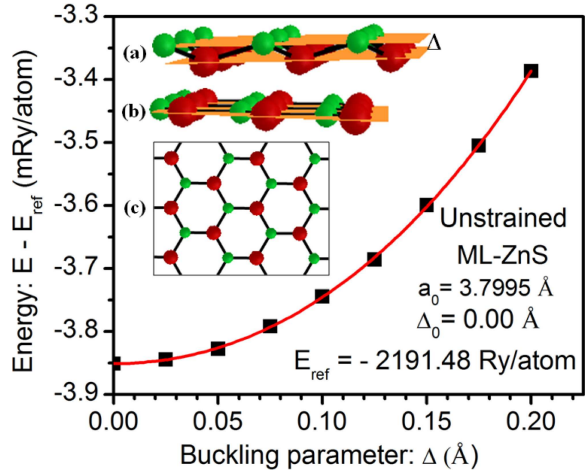}
 \caption{Buckling probe of planar (PL) ML-ZnS based on the energy
 minimization procedure with two atoms (one Zn and one S atom) per unit cell. Insets show the ball-stick model for (a) the side view of buckled (BL) ML-ZnS, (b) the side view of planar ML-ZnS and (c) the top-down view of both PL and BL ML-ZnS. In BL ML-ZnS, Zn atoms
  (large balls, red) and S atoms (small balls, green) are in two different
   parallel planes; buckling parameter is the perpendicular distance
   between those parallel planes and $\Delta = 0.00 $\AA \, for PL ML-ZnS.
   $\Delta _0$ is the value of $\Delta$ corresponding to the minimum energy at $a_0 = 3.7995$ \AA\, and E$_{\mbox{ref}}$ is a constant value of energy (-2191.48 Ry/atom) subtracted from the total energy E simply to avoid the multi-digit numbers appearing on the Y-axis.}
\end{figure}
\section{Computational Methods}
\noindent
The calculations have been performed by using the DFT based full-potential
(linearized) augmented plane wave plus local orbital (FP-(L)APW+lo) method
\cite{10} as implemented in the elk-code \cite{11}. We use the
Perdew-Zunger variant of LDA \cite{12}, the accuracy of which has been
 successfully tested in our previous studies \cite{13, 14, 15, 16, 17, 18, 19, 20, 21}
 of graphene and some graphene-like 2D crystals. For instance, our LDA value of the Fermi velocity in graphene, viz., $8.327\times 10^5$ m/s \cite{21}, matches well with that calculated using the WIEN2k code, viz., $8.2\times10^5$ m/s (LAPW$+$LDA) \cite{22} and $8.33\times10^5$ m/s (LAPW$+$GGA) \cite{23}. For plane-wave expansion in the interstitial region, we have chosen
  $|{\bf G}+{\bf k}|_{\mbox{max}}\times{\mbox{R}}_{\mbox{mt}}= 8$ (R$_{mt}$
 is the smallest muffin-tin radius in the unit cell) for deciding the
 plane-wave cut-off. The Monkhorst-Pack \cite{24} {\bf k}-point grid size
 of $20\times 20\times1$ was chosen for structural and of $30\times30\times1$
 for band structure and density of states (DOS) calculations. The total
 energy was converged within $2\mu$eV/atom. We simulate the 2D-hexagonal
 structure of ML-ZnS as a 3D-hexagonal supercell with a large value of
 $c$-parameter ($= |{\bf c}| = 40$ a.u.) and with two atoms per unit supercell. The application of homogeneous in-plane biaxial strain  $\delta < 6\%$ was simulated by varying the
 in-plane lattice parameter $a\,(=|{\bf a}| = |{\bf b}|); \delta = [(a - a_0)/a_0]\times100 \%$,
 where $a_0$ is the ground state in-plane lattice constant. For
 structure optimization, we have considered two structures of ML-ZnS as
  depicted in the insets of figure 1: (a) Buckled (BL) structure side view,
  (b) Planar (PL) structure side view and (c) top-down views of both PL and
  BL ML-ZnS in ball-stick model.

\begin{figure}
\centering
\includegraphics[scale=0.9]{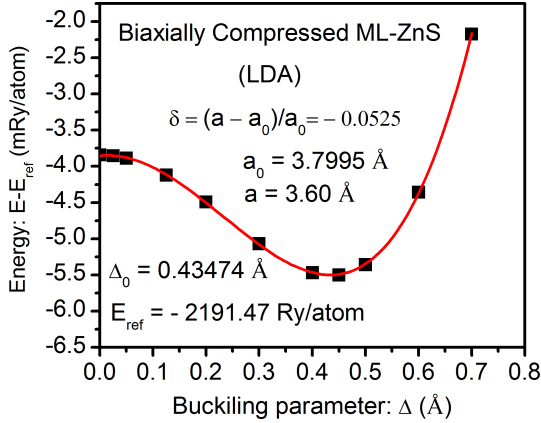}
 \caption{Buckling probe of ML-ZnS at $a = 3.60$ \AA. $\Delta _0$ is the value of $\Delta$ corresponding to the minimum energy at $a = 3.60$ \AA\, and E$_{\mbox{ref}}$ is a constant value of energy (-2191.47 Ry/atom) subtracted from the total energy E simply to avoid the multi-digit numbers appearing on the Y-axis. }
\end{figure}
\begin{figure}
\centering
\includegraphics[scale=0.9]{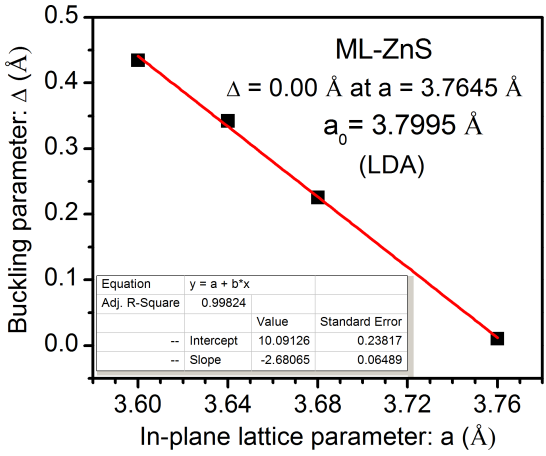}
 \caption{Variation of buckling parameter of ML-ZnS with in-plane lattice parameter.}
\end{figure}

\section{Results and Discussions}
For an assumed flat ML-ZnS, our calculated LDA value of $a_0 = 3.7995$ \AA,\,corresponding to the Zn-S bond length $d_{ZnS} = a_0/\sqrt{3}= 2.194$ \AA,\,is in agreement with theoretical GGA value of $d_{ZnS} = 2.246$ \AA \, for ZnS flat single sheet \cite{9} considering the fact that GGA usually overestimates the lattice constant. Our assumption on the flat ML-ZnS structure was tested as correct by the results depicted in Figure 1 calculated using the principle of minimum energy for the stable structure. This prediction of a graphene-like flat stable structure for ML-ZnS differs from that of ref. \cite{9} which reported that a highly reconstructed surface rather than a flat surface is the most stable structure for ML-ZnS and we attribute this difference of results to our different method of study. Having established the graphene-like planar structure of unstrained ML-ZnS, we investigated effect of homogeneous biaxial strain (both compressive and tensile strains) on the structure of ML-ZnS by calculating the buckling parameter at different values of $a > a_0$ and $a < a_0$ following the energy minimization procedure (but, now allowing for buckling) as in Figure 2. As expected, we found no buckling in ML-ZnS for tensile strains. However, for compressive strain beyond certain limit, buckling was found in ML-ZnS and at compressive strain of $5.25\%$ (corresponding to $a = 3.60$ \AA) the buckling parameter was estimated at $\Delta(a = 3.60Å) \approx 0.435$ \AA\, (Figure 2). Our calculated values of buckling parameters at four different values of $a$ (= 3.60 \AA, 3.64 \AA, 3.68 \AA, 3.76 \AA) are plotted in Figure 3, which shows a linear variation of $\Delta$ with $a$ (for 3.60 \AA $\leq a \leq$3.7645 \AA):
  \begin{equation}
 \Delta (a) = 10.09126 - 2.68065 \times a
 \end{equation}
  where both $\Delta$  and $a$ are in \AA. This equation gives us $\Delta = 0.00$\AA \,
  for $a = 3.7645$ \AA. Thus, for $a < 3.7645 $\AA \, or for compressive strain greater than $0.92\%$, ML-ZnS adopts a buckled structure. Eq. (1) may be used to calculate the buckling parameter corresponding to a particular value of $a$ within the interval
   (3.60 \AA, 3.76 \AA). For instance, using Eq.(1) we get $\Delta = 0.1192$ \AA \, for $a = 3.72$ \AA \, and we have used this value of $\Delta$ for the band structure calculation of BL ML-ZnS at $a = 3.72$ \AA.\\ 
    
\begin{figure}
\centering
\includegraphics[scale=0.9]{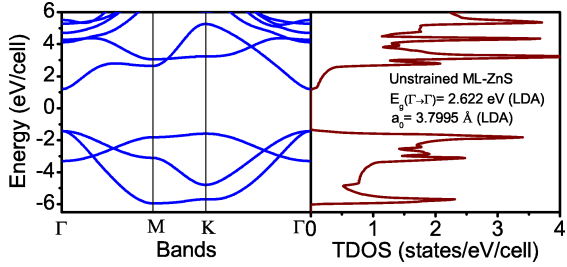}
 \caption{Bands and total density of states (TDOS) of unstrained ML-ZnS within LDA. Fermi energy $E_F$ is set at 0.00 eV.}
\end{figure}
\begin{figure}
\centering
\includegraphics[scale=1.0]{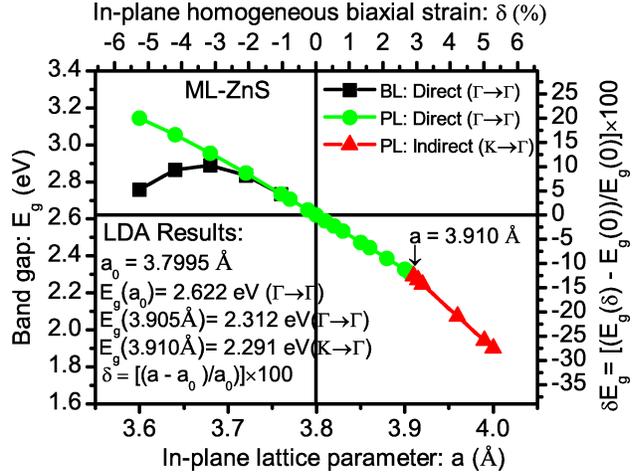}
 \caption{Variation of LDA value of band gap E$_g$ with in-plane
 lattice parameter $a$ of ML-ZnS in both planar (PL) and optimized buckled (BL) structure. The right Y-axis shows the relative variation $\delta {\mbox{E}}_g(\%)$ of E$_g$ (with respect to the unstrained value of E$_g(0) = 2.622$ eV) with in-plane homogeneous biaxial strain $\delta (\%)$ shown in the upper X-axis.}
\end{figure}
\noindent
The calculated 2D bulk modulus for the planar ML-ZnS,
$B_{\mbox{2D}}(PL)=A\left(\partial^ 2E/\partial A^2\right)_{A_{min}}=53.94 \hspace{0.1cm} \mbox{N/m}$ ($A$ is the area of the periodic cell of the 2D lattice and $A_{min}$ is the area with minimum energy)
is smaller than the corresponding calculated values of planar graphene
(214.41 N/m) and planar ML-BN (181.91 N/m) \cite{25}, reflecting the weaker
bonds in planar ML-ZnS than those in planar graphene and planar ML-BN. This
clearly means that planar ML-ZnS would be useful in applications requiring
materials with less tougher behavior than planar graphene and planar ML-BN.
Further since experimentally it is observed \cite{26} that graphene can sustain elastic tensile strain in excess of 20\%, ML-ZnS (with its 2D bulk modulus value at about one fourth of the corresponding value for graphene) is expected to sustain elastic strain about 5\% without breaking.
\\
\noindent
 The band structure and total DOS (TDOS) plot (figure 4) show that the
 unstrained ML-ZnS is a direct band gap (E$_g = 2.622$ eV, LDA value)
 semiconductor with both valence band maximum (VBM) and conduction band
 minimum (CBM) located at the $\Gamma$ point of the hexagonal Brillouin Zone (BZ).
 However, the actual band gap is expected to be larger since LDA
 underestimates the gap. Our calculated LDA band gap of 2.622 eV is
 larger than the reported GGA value of E$_g = 2.07$ eV for ZnS single
 sheet \cite{9}. For the purpose of comparison with ML-ZnO,
 we would like to note that the band gap of ML-ZnS is larger than the
 reported GGA value \cite{27} of E$_g = 1.68$ eV for ML-ZnO, which is
 identical with our calculated LDA value for ML-ZnO
 \cite{17}. This clearly suggests that ML-ZnS will be useful in device applications
  where materials with higher band gap than that of ML-ZnO is required.
 It is to be noted that both LDA and GGA are powerful methods
 to predict a correct trend in variation of the band gap \cite{28, 29, 30, 31},
  although both of them do not yield the correct band gap. Since we are
  interested in the trend as well as the relative variation in E$_g$ with
  strain, rather than its absolute value, we employed the computationally
  simpler and less time-consuming LDA for our present study.\\
 \indent
  The nature of variation of our calculated LDA value of band gap E$_g$ with
 in-plane lattice parameter $a$ is depicted in Figure 5 for both PL and
 optimized BL structure of ML-ZnS; the right Y-axis shows the relative change
 $\delta$E$_g$ (\%) in E$_g$ (which is expected to be less sensitive to
 the well known under-estimation of E$_g$ due the methods such as the
 LDA or GGA):
\begin{equation}
\delta{\mbox E}_g = \left[\left({\mbox E}_g(\delta) -
{\mbox E}_g(0)\right)/{\mbox E}_g(0)\right]\times 100 \%
\end{equation}
(with respect to the unstrained value E$_g(0) = 2.662$ eV) with in-plane
homogeneous biaxial strain $\delta = [(a - a_0)/a_0]\times 100 \%$ shown in the upper X-axis. As seen in Figure 5, a transition point from direct-to-indirect gap-phase
exists for an `$a$' value lying in the interval (3.905 \AA, 3.910 \AA) which corresponds to the a tensile strain value lying in the interval ($2.777\%,\, 2.908\%$). We would like to note that in the indirect gap-phase VBM is at the K and CBM is at the $\Gamma$ point of the BZ. In the range of tensile strain considered here in Figure 5, band gap decreases linearly with strain, although the rate of decrease in the indirect gap-phase is slightly more than in the direct gap-phase. In the compressive strain region of Figure 5, band gap variation with strain is non-linear with no change in gap-phase. For compressive strain values up to about $3\%$, band gap increases non-linearly with strain for the optimized BL structure; band gap becomes $10\%$ more than its unstrained value at compressive strain of $3\%$. However, beyond $3\%$ compressive strain, band gap shows a non-linear decrease with strain. As seen in compressive strain domain of Figure 5, our hypothetical PL structure of ML-ZnS (which is not correct beyond $0.92\% $ compressive strain) has larger band gap than the optimized BL structure of ML-ZnS.
\section{Conclusion} In conclusion, using full-potential ab initio DFT calculations we have
investigated the effect of in-plane homogeneous biaxial strain on the structural and electronic properties of ML-ZnS which is predicted to adopt a planar graphene-like honeycomb structure. We found that small compressive in-plane homogeneous biaxial strain in excess of $0.92\%$ could make ML-ZnS buckled; at compressive strain of $5.25\%$ buckling parameter as large as 0.435 \AA\, was calculated. Within compressive strain of $5.25\%$, buckling parameter was found to vary linearly with in-plane homogeneous biaxial strain. The direct band gap of ML-ZnS was found to be significantly tunable by strain-engineering. A transition point from direct-to-indirect gap-phase was predicted to exist for a tensile strain value lying within the interval $(2.777\%,\, 2.908\%)$. Within our consideration of strains less than $6\%$, the band gap decreases linearly with tensile strain in both gap-phases although the rate of decrease is a little more in the indirect gap-phase; band gap varies non-linearly (first increases and then decreases) with compressive strain with no
change in the nature of the direct band-gap. Thus, our study predicts some interesting strain-engineered structural and band-gap change of ML-ZnS. We hope, with the advancement of fabrication techniques, these predictions are testable in near future for potential applications in a variety of novel nano-devices such as strain sensors, mechatronics/straintronics, NEMS and NOMS.
      
\section*{Acknowledgement}     
{The authors acknowledge the use of the High Performance Computing Facility (Nebula-Cluster) of IIT Bombay for all the calculations in this paper.}
      

\bibliographystyle{elsarticle-num}


\end{document}